\documentclass[11pt]{article}
\usepackage{amscd}
\usepackage{amssymb}
\usepackage{amsfonts}
\usepackage{amsmath}
\usepackage{mathrsfs}
\begin{document}

\setcounter{page}{1}
\vfil

\pagestyle{plain}

\begin{center}
{\LARGE {\bf Cancellation of anomalies in a path integral \\ formulation for classical field theories}}

\bigskip
\bigskip

{\large D. Mauro}

             Department of Theoretical Physics, 
	     University of Trieste, \\
	     Strada Costiera 11, Miramare-Grignano, 34014 Trieste, Italy,\\
	     and INFN, Trieste, Italy\\
	     e-mail: {\it mauro@ts.infn.it}

\end{center}

\bigskip

\begin{abstract}
Some symmetries can be broken in the quantization process (anomalies) and
this breaking is signalled by a non-invariance of the quantum path integral measure.
In this talk we show that it is possible to formulate also classical field theories via path integral techniques. The associated classical functional measure is larger than the quantum one, because it includes some auxiliary fields. For a fermion coupled with a gauge field we prove that the way these auxiliary fields transform compensates exactly the Jacobian which arises from the transformation of the fields appearing in the quantum measure. This cancels the quantum anomaly and restores the symmetry at the classical level.
\end{abstract}

\section{Introduction}     


It is well-known that in field theory some symmetries could be broken in the quantization process. This is the phenomenon of anomalies and it is signalled by the non-invariance of the quantum functional measure under the symmetry transformations. In Ref. \cite{ennio} it is shown that it is possible to formulate also classical mechanics via path integrals. We will indicate this formalism with CPI, for Classical Path Integral. This formulation has been extended to bosonic field theories in \cite{1bis}. In this talk we will review Ref. \cite{ale} where this approach has been extended also to a theory of fermions coupled with a gauge field which, at the quantum level, present a chiral anomaly. As anomalies are not present at the classical level there must be a mechanism that cancels the anomaly in the CPI. In Ref. \cite{ale} this mechanism has been linked with the presence of several auxiliary fields in the CPI. These fields generate a Jacobian which compensates the one arising from the basic fields of the theory. This paper is organized as follows:
in Sec. 2 we will give a brief review of the CPI which was presented at this conference by Ennio Gozzi. In Sec. 3 we will extend the formalism of the CPI from the point particle to a field theory of fermions. In Sec. 4 we will briefly review the analysis of the chiral anomaly in the quantum path integral approach {\it \`a la} Fujikawa \cite{fuji}: the breaking of the chiral symmetry arises because the functional measure is not invariant under the chiral transformations. In Sec. 5 we will implement the chiral symmetry for Fujikawa's models at the CPI level. The associated functional measure turns out to have a part identical to the quantum one and another part which involves a functional integral over the auxiliary fields which appear in the CPI. These auxiliary fields play a crucial role in compensating the anomaly coming from the quantum part of the functional measure and in restoring the chiral symmetry at the classical level. 

\section{Brief review of the classical path integral}

In this section we will limit ourselves to review those features of the formalism which are crucial in order to understand the extension to the field theories. For further details we refer the interested reader to the original papers \cite{ennio}. 

It is well-known from Koopman and von Neumann's work \cite{koopman} that it is possible to formulate classical mechanics using states and operators defined in a suitable Hilbert space on phase space whose coordinates will be indicated with $\varphi\equiv (q,p)$. The main idea of this approach is to replace the probability densities $\rho(\varphi)$ with the states of a Hilbert space $\psi(\varphi)$, whose modulus square reproduces just the probability density of finding the system in a certain point of the phase space, i.e. $|\psi(\varphi)|^2=\rho(\varphi)$. The evolution of these ``Koopman-von Neumann waves" is given by the so called Liouville equation:
\begin{equation}
\displaystyle i\frac{\partial \psi}{\partial t}=\hat{L}\psi, \label{liouv}
\end{equation}
where the Liouvillian $\hat{L}$ can be written in terms of the Hamiltonian $H(\varphi)$ and of the antisymmetric matrix $\omega^{ab}=\left(\begin{array}{cc} 0 & 1 \\  -1 & 0 \end{array}\right)$ as follows: $\hat{L}=i\partial_aH\omega^{ab}\partial_b$. Because of the particular form of the operator of evolution, which is first order in the derivatives with respect to $q$ and $p$, it is easy to prove that also the probability densities $\rho(\varphi)$ evolve with the Liouville equation (\ref{liouv}). Now, since every theory formulated with operatorial techniques can be rewritten in the path integral language, it must be possible to reformulate also classical mechanics via path integrals. This has been done in Ref. \cite{ennio} by starting from the following question: which is the probability density of going from the point $\varphi_i$ of the phase space at time $t_i$ to the point $\varphi_f$ at time $t_f$? In classical mechanics we have only two possibilities: this probability is one if the point $\varphi_f$ lies at time $t_f$ on the classical path $\phi^a_{\textrm{\scriptsize{cl}}}(t;\varphi_i)$ and zero otherwise. By classical path $\phi^a_{\textrm{\scriptsize{cl}}}(t;\varphi_i)$ we mean the path which solves the classical Hamilton's equations of motion $\dot{\varphi}^a=\omega^{ab}\partial_bH(\varphi)$ with the initial condition $\varphi(t_i)=\varphi_i$.This result can be written as a path integral
over $\varphi$ of a functional Dirac delta which gives weight one only to the classical path associated with the initial conditions $\varphi_i$. 
\begin{equation}
\displaystyle Z=\langle \varphi_f;t_f|\varphi_i;t_i\rangle =\int {\cal D}^{\prime\prime} \varphi \,
\delta\left[\varphi^a-\phi^a_{\textrm{\scriptsize{cl}}}(t;\varphi_i)\right]. \label{prob}
\end{equation}
The double prime on ${\cal D}$ means that the initial and the final point in the phase space are fixed. 
If we now use the properties of the Dirac deltas we can replace the Dirac delta on the solutions of the equations of motion with a Dirac delta on the equations of motion plus a functional determinant:
\begin{equation}
\displaystyle \delta\left[\varphi^a-\phi^a_{\textrm{\scriptsize{cl}}}(t;\varphi_i)\right]=\delta(\dot{\varphi}^a-\omega^{ab}\partial_bH)  \,\textrm{det}(\partial_t\delta_c^a-\omega^{ab}\partial_b\partial_cH).  \label{deter}
\end{equation}
We can use the Fourier representation of the Dirac delta to exponentiate the Dirac delta on the equations of motion via a functional integral over an auxiliary variable $\lambda_a$. Furthermore we can use the Faddeev-Popov trick to exponentiate the determinant in the RHS of (\ref{deter}) via a couple of Grassmannian odd variables $c^a$ and $\bar{c}_a$. The final result is that the probability amplitude (\ref{prob}) can be rewritten as the following path integral:
\begin{equation}
\displaystyle Z=\int {\cal D}^{\prime\prime}\varphi{\cal D}\lambda{\cal D}c{\cal D}\bar{c}
\; \exp i \int_{t_i}^{t_f} {\textrm{d}}t \, \widetilde{\cal L} \label{cpi}
\end{equation}
where the functional integral is extended not only over the phase space variables $\varphi$ but also over all the auxiliary variables $\lambda, c$ and $\bar{c}$.
The Lagrangian $\widetilde{\cal L}$ is given by
\begin{displaymath}
\displaystyle \widetilde{\cal L}=\lambda_a\dot{\varphi}^a+i\bar{c}_a\dot{c}^a-{\cal H},
\qquad {\cal H}=\lambda_a\omega^{ab}\partial_bH+i\bar{c}_a\omega^{ab}\partial_b\partial_dHc^d.
\end{displaymath}
Let us now define the commutators as Feynman did for quantum mechanics, i.e. using the following rule
\begin{displaymath}
\displaystyle \langle \left[O_1(t),O_2(t)\right]\rangle\equiv \lim_{\epsilon \to 0}
\langle O_1(t+\epsilon)O_2(t)\pm O_2(t+\epsilon)O_1(t)\rangle.
\end{displaymath}
We get that $[\hat{\varphi}^a,\hat{\varphi}^b]=0$, i.e. the position $\hat{q}$ and the momentum $\hat{p}$ commute, which confirms that we are doing classical and not quantum mechanics. The only non-zero graded commutators are given by:
\begin{displaymath}
\displaystyle [ \hat{\varphi}^a,\hat{\lambda}_b]_{\scriptstyle{-}}=i\delta_b^a,
\qquad [ \hat{c}^a,\hat{\bar{c}}_b]_{\scriptstyle{+}}=\delta_b^a.
\end{displaymath}
The previous commutators can be realized by taking $\hat{\varphi}^a$ and $\hat{c}^a$ as multiplicative operators and $\hat{\lambda}_a$ and $\hat{\bar{c}}_a$ as the following derivative operators:
\begin{displaymath}
\displaystyle \hat{\lambda}_a=-i\frac{\partial}{\partial \varphi^a}, \qquad \hat{\bar{c}}_a=\frac{\partial}{\partial c^a}.
\end{displaymath}
Via this choice of operators the Hamiltonian which appears in the weight of the classical path integral (\ref{cpi}) becomes the following operator:
\begin{displaymath}
\displaystyle \hat{\cal H}=-i\omega^{ab}\partial_bH\partial_a-i\omega^{ab}\partial_b\partial_dHc^d \frac{\partial}{\partial c^a}.
\end{displaymath}
The first operator is just the Liouvillian which enters the equation of evolution (\ref{liouv}) of the ``Koopman-von Neumann waves". In this sense we can say that the path integral (\ref{cpi}) can be considered as the functional counterpart of the Koopman-von Neumann formalism. This path integral formulation of classical mechanics is very rich from the geometrical point of view and part of this richness has been explored in Ref. \cite{geometry}. 
As now classical mechanics has been formulated using the same tools of quantum mechanics, it is easy to make a comparison between the two theories \cite{interplay}
and to study the relative interplay.
A typical example of the interplay between classical and quantum mechanics in field theories is given by the issue of anomalies, i.e. symmetries which are present at the classical level but that are broken by the quantization procedure. For example \cite{ale} a field theory of fermions coupled with a gauge field is invariant under chiral transformations in classical mechanics but leads to a chiral anomaly at the quantum level. This talk is based just on Ref. \cite{ale} to which we refer the interested reader for further technical details.

\section{Classical path integral for fermions}

Since the main goal of this paper is to study the chiral symmetry in the framework of the CPI, the first thing that we have to do is to extend the formalism of the CPI from the point particle case, that we have briefly reviewed in Sec. 2, to the case of a field theory of fermions endowed with chiral symmetry. Let us start from the simple case of a free massless fermion theory. The Lagrangian of the system is given by: 
\begin{equation}
\displaystyle L=i\int {\textrm{d}}{\mathbf{x}}\,\bar{\psi}(x)\gamma^{\mu}
\partial_{\mu}\psi(x) \label{lagdirac}
\end{equation}
where $\psi(x)$ is a Grassmannian odd field and $\bar{\psi}(x)$ is defined as $\bar{\psi}=\psi^{\dagger}\gamma^{0}$. The Hamiltonian associated with the Lagrangian (\ref{lagdirac}) is:
\begin{displaymath}
\displaystyle H=-i\int {\textrm{d}}{\mathbf{x}} \;\psi^{\dagger}(x)\gamma^{0}
\gamma^{l}\partial_{l}\psi(x). 
\end{displaymath}
The Euler-Lagrange equations which can be derived from (\ref{lagdirac}) are:
\begin{equation}
\dot{\psi}+\gamma^{0}\gamma^{l}\partial_{l}\psi=0, 
\qquad\quad\dot{\psi}^{\dagger}+\partial_{l}\psi^{\dagger}\gamma^{0}
\gamma^{l}=0. \label{Lageq}
\end{equation}
From (\ref{lagdirac}) we have that $\psi^{\dagger}$ can be considered as the momentum canonically conjugated to $\psi$. So if we want to keep the notation as similar as possible to the one used in Sec. 2 we can collect together $\psi$ and $\psi^{\dagger}$ in a unique field $\Psi^a=(\psi, \psi^{\dagger})$. Furthermore we can introduce the following symplectic matrix 
$\omega^{ab}=\left(\begin{array}{cc} 0 & 1 \\  1 & 0 \end{array}\right)$. 
Let us notice that, differently than in the case of the point particle, the symplectic matrix is symmetric because the fields $\psi$ and $\psi^{\dagger}$ are Grassmannian odd. Using $\Psi^a$ and $\omega^{ab}$ the equations of motion (\ref{Lageq}) can be written in a compact way as 
\begin{displaymath}
\displaystyle \dot{\Psi}^a(x)=-i\omega^{ab}\frac{\partial H}{\partial \Psi^b(x)}
\end{displaymath} 
where $\partial$ must be intended as a functional derivative. 

We have now all the ingredients to implement the CPI for a free theory of massless fermions. As have seen in Sec. 2, the starting point is the definition of the following generating function for classical mechanics:
\begin{displaymath}
Z_{\scriptscriptstyle \textrm{\tiny{CM}}}[0]=\int {\mathcal D}\Psi^a\,
{\delta}[\Psi^a-\Psi^a_{\textrm{\scriptsize{cl}}}]
\end{displaymath}
where $\Psi^a_{\textrm{\scriptsize{cl}}}$ is, as usual, the solution of the classical equations of motion (\ref{Lageq}). We can then pass from the delta function of the solutions of the equations of motion to the delta of the equations of motion themselves. In this step the inverse of a functional determinant makes its appearance because of the Grassmannian nature of $\psi$: 
\begin{equation}
\displaystyle Z_{\scriptscriptstyle \textrm{\tiny{CM}}}= 
\int {\mathcal D}\Psi^a \,{\delta}\biggl(\dot{\Psi}^a+i\omega^{ab}
\frac{\partial H}{\partial \Psi^b}\biggr)\textrm{det}^{-1}\biggl[\delta^a_d \partial_t
\delta({\mathbf{x}}-{\mathbf{y}})+i\omega^{ab}\frac{\partial^2H}{\partial \Psi^d({\mathbf{y}})
\partial\Psi^b({\mathbf{x}})}\biggr]. \label{passage}
\end{equation}
For a formal point of view the steps are very similar to the ones that we have seen in the case of a point particle. Nevertheless we want to stress again the two main novelties:\\
1) the presence on the RHS of (\ref{passage}) of an inverse functional determinant instead of a functional determinant due to the fact that $\Psi^a$ is a Grassmannian odd field; \\
2) the presence of functional derivatives due to the fact that $\Psi^a$ is a field and not an ordinary variable.

The Grassmannian odd parity of the fields $\psi$ and $\psi^{\dagger}$ implies some differences also in the other steps of the CPI procedure. In fact, when we exponentiate the Dirac delta on the equations of motion and the inverse of the functional determinant we have to introduce, as usual, auxiliary fields. In this case the fields used to exponentiate the Dirac delta of the equations of motion are Grasmannian odd, let us indicate them as $\lambda_a=(\lambda_{\psi},\lambda_{\psi^{\dagger}})$. Furthermore the fields used to exponentiate the inverse of the functional determinant of Eq. (\ref{passage}) are Grassmannian even $c^a=(c^{\psi},c^{\psi^{\dagger}})$ and $\bar{c}_a=(\bar{c}_{\psi},\bar{c}_{\psi^{\dagger}})$. The final result of these manipulations is that the classical generating functional can be written as:
\begin{equation}
\displaystyle Z_{\scriptscriptstyle \textrm{\tiny{CM}}}[0]=
\int {\mathcal D}\Psi^a {\mathcal D}\lambda_a{\mathcal D}c^a{\mathcal
D}\bar{c}_a \,\textrm{exp} \biggl[
i\int {\textrm{d}}x \,\widetilde{\cal L}\biggr] \label{final}
\end{equation}
where $\widetilde{\cal L}$ is the following Lagrangian density:
\begin{eqnarray}
\displaystyle \label{suplag} \widetilde{\cal L}&=&
\lambda_{\psi}(\dot{\psi}+\gamma^{0}
\gamma^{l}\partial_{l}\psi)-(\dot{\psi}^{\dagger}+
\partial_{l}\psi^{\dagger}\gamma^{0}\gamma^{l})
\lambda_{\psi^{\dagger}}+\nonumber\\
&&+i\bar{c}_{\psi}(\dot{c}^{\psi}+\gamma^{0}
\gamma^{l}\partial_{l}c^{\psi})+
i(\dot{c}^{\psi^{\dagger}}+\partial_{l}c^{\psi^{\dagger}}\gamma^{0}
\gamma^{l})\bar{c}_{\psi^{\dagger}}. \nonumber
\end{eqnarray}

The path integral (\ref{final}) is not the only path integral for a classical field theory of massless fermions that we can implement. In fact we could have started considering $\psi$ and $\bar{\psi}$, instead of $\psi$ and $\psi^{\dagger}$, as independent fields. 
Using the equations of motion:
\begin{equation}
\dot{\psi}+\gamma^{0}\gamma^{l}\partial_{l}\psi=0\qquad\quad 
\dot{\bar{\psi}}+\partial_{l}\bar{\psi}\gamma^{l}\gamma^{0}=0 \label{eqbar}
\end{equation}
and the standard CPI procedure we would have got the following generating functional:
\begin{equation}
\displaystyle \bar{Z}_{\scriptscriptstyle \textrm{\tiny{CM}}}[0]=\int {\mathcal D}\bar{\psi}{\mathcal D}\psi{\mathcal D}\lambda_{\bar{\psi}}{\mathcal D}\lambda_{\psi}\cdots \textrm{exp}\biggl[i\int \, {\textrm{d}}x \,\bar{\cal L}\biggr] \label{pathz}
\end{equation}
where $\cdots$ indicates the functional integration over the auxiliary fields $(c^{\psi},c^{\bar{\psi}})$ and $(\bar{c}_{\psi},\bar{c}_{\bar{\psi}})$, which we must introduce to exponentiate the inverse of the functional determinant. Before going on, let us notice that in Eq. (\ref{pathz}) the first part of the functional measure, i.e. ${\mathcal D}\bar{\psi}{\mathcal D}\psi$, is identical to the functional measure of the quantum path integral. In addition, we have also the functional integration over all the auxiliary fields. The Lagrangian $\bar{\cal L}$ which appears in the weight of the path integral (\ref{pathz}) is given by 
\begin{equation}
\bar{\cal L}=\lambda_{\psi}\gamma^0\gamma^{\mu} \partial_{\mu}\psi - (\partial_{\mu}\bar{\psi})\gamma^{\mu}\gamma^0\lambda_{\bar{\psi}}+i\bar{c}_{\psi}\gamma^0\gamma^{\mu}\partial_{\mu}c^{\psi}+i(\partial_{\mu}c^{\bar{\psi}})\gamma^{\mu}\gamma^0\bar{c}_{\bar{\psi}}. \label{suplag2} 
\end{equation}
The fields $\bar{\psi}$ and $\psi^{\dagger}$ are related by the usual equation $\bar{\psi}=\psi^{\dagger}\gamma^0$. Similar relationships hold also for the auxiliary fields entering the definitions of the two CPIs of Eqs. (\ref{final}) and (\ref{pathz}):
\begin{displaymath}
\lambda_{\bar{\psi}}=
\gamma^{0}\lambda_{\psi^{\dagger}}, \qquad c^{\bar{\psi}}=c^{\psi^{\dagger}}
\gamma^{0}, \qquad \bar{c}_{\bar{\psi}}=\gamma^{0}\bar{c}_{\psi^{\dagger}}.
\end{displaymath}
If we rewrite the Lagrangian (\ref{suplag2}) as
\begin{displaymath}
\bar{\cal L}=\lambda_{\psi}\dot{\psi}-\dot{\bar{\psi}}\lambda_{\bar{\psi}}+i\bar{c}_{\psi}\dot{c}^{\psi}+i\dot{c}^{\bar{\psi}}\bar{c}_{\bar{\psi}}-\bar{\cal H}
\end{displaymath}
we can then read off the following Hamiltonian:
\begin{equation}
\bar{\cal H}=-\lambda_{\psi}\gamma^{0}\gamma^{l}\partial_{l}\psi+(\partial_{l}\bar{\psi})\gamma^{l}
\gamma^{0}\lambda_{\bar{\psi}}-i\bar{c}_{\psi}\gamma^{0}\gamma^{l}\partial_{l}c^{\psi}
-i(\partial_{l}c^{\bar{\psi}})\gamma^{l}\gamma^{0}\bar{c}_{\bar{\psi}} \label{uno}
\end{equation}
and the following non-zero anticommutators:
\begin{equation}
\Bigl[\psi_{\alpha}({\mathbf{x}},t),\lambda_{\psi, \beta}({\mathbf{y}},t)\Bigr]=
i\delta_{\alpha\beta}\delta({\mathbf{x}}-{\mathbf{y}}), \quad  \Bigl[\bar{\psi}_{\alpha}({\mathbf{x}},t),\lambda_{\bar{\psi}, \beta}({\mathbf{y}},t)\Bigr]=
i\delta_{\alpha\beta}\delta({\mathbf{x}}-{\mathbf{y}}). \label{due}
\end{equation}
Using Eqs. (\ref{uno}) and (\ref{due}) we can rewrite the equations of motion (\ref{eqbar}) for the spinors $\Psi^a$ as 
\begin{displaymath}
\displaystyle \dot{\Psi}^a=i \left[ \Psi^a, \int {\textrm{d}}{\mathbf{x}}\, \bar{\cal H}\right].
\end{displaymath}
Before going on, we want to mention a technical point that we will use later on. The space underlying the CPI is an enlarged Hilbert space. Since we are dealing with Grassmannian odd fields, we have to clarify which is the scalar product that we use. There are different possible choices but the only positive definite scalar product is the one in which the Hermitian conjugate of a Grassmannian odd field is proportional to the momentum canonically conjugated to that field \cite{deotto}, i.e.
\begin{equation}
\displaystyle -i\lambda_{\psi,\beta}=(\psi_{\beta})^{\dagger}, \quad -i\bar{\psi}_{\alpha}=(\lambda_{\bar{\psi},\alpha})^{\dagger}. \label{scpr}
\end{equation}

\section{Review of the chiral anomaly in quantum field theory}

Before studying how the chiral symmetry can be implemented at the CPI level, we want to briefly review how a chiral anomaly arises when we use the path integral technique to quantize the field theory \cite{fuji}. If we couple the massless fermion with a U(1) gauge field then the Lagrangian which describes the system is:
\begin{equation}
\displaystyle {\mathcal L}=i\bar{\psi}\gamma^{\mu}D_{\mu}\psi-\frac{1}{4}
F_{\mu \nu}F^{\mu \nu}, \label{lagfuj}
\end{equation}
where $D_{\mu}=\partial_{\mu}+ieA_{\mu}$ is the usual covariant derivative.
The Lagrangian density (\ref{lagfuj}) is symmetric under the following infinitesimal chiral transformations:
\begin{equation}
\begin{array}{l}
\displaystyle \psi({x})\,\longrightarrow \,\textrm{exp}\,[i\alpha \gamma^5]\,
\psi({x})\; \simeq \;[1+i\alpha\gamma^5 ]\psi({x}), \medskip\\
\displaystyle \bar{\psi}({x})\,\longrightarrow \, \bar{\psi}({x})\,\textrm{exp}\,
[i\alpha\gamma^5] \; \simeq \; \bar{\psi}({x})[1+i\alpha\gamma^5]. \label{chirtr}
\end{array}
\end{equation}
The Noether current associated with this symmetry is $J^{\mu}_5=\bar{\psi}\gamma^{\mu}\gamma^5\psi$. When we quantize the theory the Lagrangian density (\ref{lagfuj}) appears in the weight of the following path integral:
\begin{displaymath}
\displaystyle Z_{\scriptscriptstyle \textrm{\tiny{QM}}}[0]=
\int [{\mathcal D}A_{\mu}(x)]{\mathcal D}\bar{\psi}(x)
{\mathcal D}\psi(x)\,\textrm{exp} \,\biggl[\frac{i}{\hbar}\int {\textrm{d}}x \,{\mathcal L}\biggr],
\end{displaymath} 
where $[{\mathcal D}A_{\mu}(x)]$ is the part of the functional measure necessary to describe the gauge field. 
Under local chiral transformations the Lagrangian density changes as follows
\begin{displaymath}
{\mathcal L} \; \longrightarrow \; {\mathcal L} -\partial_{\mu}\alpha(x)J_5^{\mu}(x).
\end{displaymath}
If the functional measure were invariant under the chiral transformations (\ref{chirtr})  then the requirement that the generating functional does not depend on the parameter $\alpha$, i.e. 
\begin{displaymath}
\displaystyle \frac{\partial Z_{\scriptscriptstyle \textrm{\tiny{QM}}}}{\partial \alpha(x)}\biggl|_{\alpha=0}=0,
\end{displaymath}
would imply that $\langle \partial_{\mu}J_5^{\mu}\rangle=0$, i.e. the chiral symmetry would be present also at the quantum level. Unfortunately, as Fujikawa pointed out at the end of the 1970s \cite{fuji}, the functional measure of the quantum path integral is not invariant under the chiral transformations. In other words, if we perform the chiral transformations (\ref{chirtr}), then the functional measure ${\mathcal {D}}\bar{\psi}{\mathcal D}\psi$ generates a Jacobian $J$ different from zero:
\begin{displaymath}
\displaystyle {\mathcal D}\bar{\psi}^{\prime}{\mathcal D}\psi^{\prime}=
J\,{\mathcal D}\bar{\psi}{\mathcal D}\psi\equiv \exp \biggl[ i\int {\textrm{d}}x\,\alpha(x){\mathcal A}[A_{\mu}](x)\biggr]\,
{\mathcal D}\bar{\psi}{\mathcal D}\psi. 
\end{displaymath}
This non-invariance of the functional measure implies a breaking of the chiral symmetry. In fact the requirement that the generating functional $Z_{\scriptscriptstyle \textrm{\tiny{QM}}}[0]$ does not depend on $\alpha$ implies that the mean value of $\partial_{\mu}J^{\mu}_5$ becomes:  
\begin{displaymath}
\langle \partial_{\mu}J_5^{\mu}\rangle =\hbar \langle {\mathcal A}\rangle. 
\end{displaymath}
To evaluate explicitly ${\mathcal A}$ or, equivalently, the Jacobian $J$ associated with the chiral transformation of the functional measure, Fujikawa \cite{fuji} considered a complete set of eigenstates $\{\phi_n(x)\}$ of a Hermitian operator and expanded the fields $\psi$ and $\bar{\psi}$ as follows:
\begin{displaymath}
\displaystyle \psi({x})=\sum_nb_n\phi_n({x}), \qquad
\bar{\psi}({x})=\sum_n\phi_n^{\dagger}({x})\bar{b}_n.
\end{displaymath}
As a consequence, the functional measure of the quantum path integral can be rewritten as a product of standard integrals over the coefficients of the expansion $\displaystyle {\mathcal D}\bar{\psi}{\mathcal D}\psi=\prod_n {\textrm{d}}\bar{b}_n{\textrm{d}}b_n$ and the Jacobian $J$ can be rewritten as $J=[\textrm{det} \, C]^{-2}$ where the matrix $C$ is given by 
\begin{equation}
\displaystyle C_{mn}\equiv \delta_{mn}+i\int {\textrm{d}}x\, \alpha({x})\phi_m^{\dagger}({x})
\gamma^5\phi_n({x}). \label{matrixc}
\end{equation}
The determinant of $C$ is an ill-defined quantity which has to be regularized. Using a gauge invariant regularization one gets that the determinant of $C$ is different from zero and so a chiral anomaly appears in the quantum theory.

\section{Chiral symmetry and the classical path integral}

In Sec. 3 we have implemented the CPI for a classical field theory of massless fermions.
If we include also the interaction with an external gauge field then the CPI Lagrangian which describes the fermions can be obtained from the one of Eq. (\ref{suplag2}) replacing the standard derivatives $\partial_{\mu}$ with the covariant ones $D_{\mu}$:
\begin{displaymath}
\bar{\cal L}=\lambda_{\psi}\gamma^0\gamma^{\mu}D_{\mu}\psi-(D_{\mu}\bar{\psi})\gamma^{\mu}\gamma^0\lambda_{\bar{\psi}}+i\bar{c}_{\psi}\gamma^0\gamma^{\mu}D_{\mu}c^{\psi}+i(D_{\mu}c^{\bar{\psi}})\gamma^{\mu}\gamma^0\bar{c}_{\bar{\psi}}.
\end{displaymath}
This Lagrangian is invariant under the following infinitesimal transformations:
\begin{eqnarray}
\label{trapsi} \delta \psi=i\alpha \gamma^5\psi, &&\quad \delta \bar{\psi}=
i\alpha\bar{\psi}\gamma^5,\nonumber\\
\delta \lambda_{\psi}=-i\alpha \lambda_{\psi}\gamma^5, &&\quad \delta \lambda_{\bar{\psi}}
=-i\alpha \gamma^5 \lambda_{\bar{\psi}},\\
\delta c^{\psi}=i\alpha \gamma^5c^{\psi}, &&\quad \delta c^{\bar{\psi}}=i\alpha c^{\bar{\psi}}\gamma^5, 
\nonumber\\
\delta \bar{c}_{\psi}=-i\alpha \bar{c}_{\psi}\gamma^5, &&\quad
\delta\bar{c}_{\bar{\psi}}=-i\alpha \gamma^5\bar{c}_{\bar{\psi}}. \nonumber 
\end{eqnarray}
The transformations in the first line of (\ref{trapsi}) are just the standard chiral transformations of spinors, see Eq. (\ref{chirtr}). The other transformations implement the chiral symmetry at the level of the auxiliary fields of the CPI. The associated Noether current is given by
\begin{displaymath}
\widetilde{J}^{\mu}_5=-i\biggl[\lambda_{\psi}\gamma^{0}\gamma^{\mu}\gamma^{5}\psi-
\bar{\psi}\gamma^{5}
\gamma^{\mu}\gamma^{0}\lambda_{\bar{\psi}}+i\bar{c}_{\psi}\gamma^{0}\gamma^{\mu}\gamma^{5}c^{\psi}
+ic^{\bar{\psi}}\gamma^{5}\gamma^{\mu}\gamma^{0}\bar{c}_{\bar{\psi}}\biggr]
\end{displaymath}
and leads to the following conservation law: $\partial_{\mu}\widetilde{J}^{\mu}_5=0$. The Lagrangian $\bar{\cal L}$, which appears in the weight of the CPI, is invariant under the chiral transformations, but in the CPI there is also the functional measure and in Fujikawa's approach it is just the non-invariance of the functional measure which produces the chiral anomaly at the quantum level. So let us consider again the 
generating functional for classical mechanics:
\begin{equation}
\displaystyle \bar{Z}_{\scriptscriptstyle \textrm{\tiny{CM}}}[0]=\int {\mathcal D}\bar{\psi}{\cal D}\psi{\cal D}\lambda_{\bar{\psi}}{\cal D}\lambda_{\psi} \cdots \exp \left[ i\int \textrm{d}x\, \bar{\cal L}\right]. \label{cgf}
\end{equation}
Since we are dealing with classical mechanics we expect no anomaly which means that the functional measure of the CPI (\ref{cgf}) should be invariant under the chiral transformations (\ref{trapsi}).
Nevertheless we know that ${\cal D}\bar{\psi}{\cal D}\psi$ is just the functional measure of the quantum path integral and such a measure is not invariant under the chiral transformations. The measure ${\cal D}\bar{\psi}{\cal D}\psi$ appears also in the classical path integral (\ref{cgf}). As a consequence, to have a symmetry at the classical level we must compensate this non-invariance and this can only happen if also the functional measure over the auxiliary fields of the CPI is not invariant under chiral transformations. In particular, it should transform under chiral transformations in such a way to produce a Jacobian which cancels the contribution coming from the quantum part of the measure, i.e. from ${\cal D}\psi{\cal D}\psi^{\dagger}$. 

Let us use Fujikawa's techniques, that we have reviewed in the previous section, to prove that the whole functional measure of the CPI is invariant under chiral transformations. To do this, let us remember the definition (\ref{scpr}) of the scalar product in the enlarged space of the CPI. According to this scalar product, if we expand $\psi$ over an orthonormal set of eigenstates $\phi_n(x)$ then we can expand the canonically conjugated momentum
$\lambda_{\psi}(x)$ over the basis given by $\phi_n^{\dagger}(x)$:
\begin{equation}
\displaystyle \psi(x)=\sum_nb_n\phi_n(x), \qquad \lambda_{\psi}(x)=\sum_n\phi_n^{\dagger}(x)\beta_n. \label{exp}
\end{equation}
From Eq. (\ref{trapsi}) we have that $\psi$ and $\lambda_{\psi}$ transform under chiral transformations with a different sign. This immediately implies that also the coefficients of their expansion (\ref{exp}) transform in a different way:
\begin{eqnarray}
&&\displaystyle \psi^{\prime}(x)=\left[ 1+i\alpha(x)\gamma^5\right]\psi(x) \, \Longrightarrow \, b_m^{\prime}=C_{mn}b_n \bigskip \nonumber \\
&&\displaystyle \lambda_{\psi}^{\prime}(x)=\lambda_{\psi}(x)\left[1-i\alpha(x)\gamma^5\right] \, \Longrightarrow \, \beta_m^{\prime}=\beta_mD_{mn}
\nonumber
\end{eqnarray}
where $C$ and $D$ are the following functional matrices: 
\begin{equation}
\begin{array}{l}
\displaystyle C_{mn}\equiv \delta_{mn}+i\int {\textrm{d}}x \; \alpha(x)\phi_m^{\dagger}(x)\gamma^5\phi_n(x) \smallskip  \\  
\displaystyle D_{mn}\equiv \delta_{mn}-i\int {\textrm{d}}{x}\; \alpha({x}) \phi_m^{\dagger}({x})
\gamma^5\phi_n({x}). \label{cidi}
\end{array}
\end{equation}
Analogously we can consider the expansions of $\bar{\psi}(x)$ and $\lambda_{\bar{\psi}}(x)$
\begin{displaymath}
\bar{\psi}(x)=\sum_m\phi_m^{\dagger}(x)\bar{b}_m,  \qquad\quad
\displaystyle \lambda_{\bar{\psi}}(x)=\sum_m\bar{\beta}_m\phi_m(x). 
\end{displaymath}
Also in this case the chiral transformations of $\bar{\psi}$ and $\lambda_{\bar{\psi}}$ differ by a sign. 
The chiral transformations of the coefficients $\bar{b}$ and $\bar{\beta}$ can be easily derived and they involve the usual matrices $C$ and $D$ of Eq. (\ref{cidi}):
\begin{displaymath}
\bar{b}_n^{\prime}=\sum_m\bar{b}_mC_{mn},\qquad \quad \bar{\beta}_m^{\prime}=\sum_nD_{mn}\bar{\beta}_{n}.
\end{displaymath}
If we collect together all the results above we have that the functional measure of the CPI can be rewritten as a product of ordinary integrals over the coefficients $b, \bar{b}, \beta,\bar{\beta}$:
\begin{displaymath}
\displaystyle {\mathcal D}\psi{\mathcal D}\bar{\psi}{\mathcal D}\lambda_{\psi}
{\mathcal D} \lambda_{\bar{\psi}}=\prod_m {\textrm{d}}b_m{\textrm{d}}\bar{b}_m{\textrm{d}}\beta_m{\textrm{d}}\bar{\beta}_m.
\end{displaymath}
Under a chiral transformation it transforms as follows:
\begin{equation}
\displaystyle {\mathcal D}\psi^{\prime}{\mathcal D}\bar{\psi}^{\prime}
{\mathcal D}\lambda_{\psi}^{\prime}{\mathcal
D}\lambda_{\bar{\psi}}^{\prime}=\widetilde{J} \,
\cdot {\mathcal D}\psi{\mathcal D}\bar{\psi}{\mathcal D}\lambda_{\psi}{\mathcal
D}\lambda_{\bar{\psi}}, \label{supjac}
\end{equation}
i.e. via a Jacobian $\widetilde{J}$ which is given by: $\widetilde{J}=[\textrm{det} \, C]^{-2}\cdot [\textrm{det} \, D]^{-2}$. As we have seen in the previous section the term $[\textrm{det}\, C]^{-2}$ is just the Jacobian coming from the chiral transformations of the quantum part of the functional measure. In addition there appears also the term $[\textrm{det}\, D]^{-2}$ which comes from the transformations of the auxiliary fields $\lambda$. The two matrices $C$ and $D$ are one the inverse of the other, in fact:
\begin{eqnarray}
\displaystyle (CD)_{nl}&=&\sum_m \Bigl(\delta_{nm}+i\int {\textrm{d}}x \; \alpha(x)
\phi_n^{\dagger}(x)\gamma^5\phi_m(x)\Bigr) \cdot\nonumber\\
&& \cdot \Bigl(\delta_{ml}-i\int {\textrm{d}}x \; \alpha(x) \phi_m^{\dagger}(x)\gamma^5\phi_l(x)\Bigr)=
\delta_{nl}+O(\alpha^2). \nonumber
\end{eqnarray}
This immediately means that $\widetilde{J}=1$, which implies from (\ref{supjac}) that the functional measure ${\mathcal D}\psi{\mathcal D}\bar{\psi}{\mathcal D}\lambda_{\psi}{\mathcal D}\lambda_{\bar{\psi}}$ is invariant under chiral transformations . So we can say that the contribution to the Jacobian $\widetilde{J}$ coming from the functional measure over the auxiliary fields ${\cal D}\lambda_{\psi}{\cal D}\lambda_{\bar{\psi}}$ compensates exactly the contribution to the Jacobian coming from ${\cal D}\psi{\cal D}\bar{\psi}$ and it produces a cancellation of the quantum anomaly. Since a similar argument holds for the functional measure ${\mathcal D}c^{\psi}{\mathcal D}c^{\bar{\psi}}{\mathcal D}\bar{c}_{\psi}{\mathcal D}\bar{c}_{\bar{\psi}}$ we can conclude that the entire CPI functional measure is invariant under chiral transformations and no chiral anomaly arises.

What we would like to do next is to extend this analysis to the case of scale anomaly both in quantum mechanical models and in field theories. In particular we would like to understand whether the auxiliary fields of the CPI play also in this case a role in cancelling the quantum scale anomaly and in restoring the scale symmetry at the classical level.

\bigskip
{\small This work has been supported by grants from the University of Trieste, MIUR and INFN of Italy. I would like to thank E. Gozzi and A. Silvestri for the essential role they had in this project.}

\end{document}